\documentclass[preprint2]{aastex}

\usepackage{flushend} %% balance columns on last page
\usepackage{color}
\newcommand{\be}{\begin{equation}}
\newcommand{\ee}{\end{equation}}

\newcommand{\cC}{\mathcal C}

\definecolor{Dred}{rgb}{0.312,0.070,0.070}
\definecolor{Dblue}{rgb}{0.070,0.070,0.312}
\definecolor{Dgreen}{rgb}{0.070,0.312,0.070}
\definecolor{Db}{rgb}    {0.050,0.0,0.320}

\newcounter{note}
\let\oldmarginpar\marginpar
\renewcommand\marginpar[1]{\-\oldmarginpar[\raggedleft\footnotesize #1]{\raggedright\footnotesize #1}}

\shorttitle{Compact Radio Jets on Sub-parsec Scales}
\shortauthors{Lee et al.}

\begin{document}

\title{Acceleration of Compact Radio Jets on Sub-parsec Scales}

\author{
Sang-Sung Lee\altaffilmark{1,2},      %  sslee@kasi.re.kr 
Andrei P. Lobanov\altaffilmark{3},     %  alobanov@mpifr-bonn.mpg.de 
Thomas P. Krichbaum\altaffilmark{3},   %  tkrichbaum@mpifr-bonn.mpg.de
and
J. Anton Zensus\altaffilmark{3}       %  azensus@mpifr-bonn.mpg.de
}

\altaffiltext{1}{Korea Astronomy and Space Science Institute, 
776 Daedeok-daero, Yuseong-gu, Daejeon 34055,
Republic of Korea; sslee@kasi.re.kr}
\altaffiltext{2}{Korea University of Science and Technology,
217 Gajeong-ro, Yuseong-gu, Daejeon 34113,
Republic of Korea}
\altaffiltext{3}{Max-Planck-Institut f\"ur Radioastronomie,
Auf dem H\"ugel 69, 53121 Bonn,
Germany; alobanov, tkrichbaum, azensus@mpifr-bonn.mpg.de}

\begin{abstract}
Jets of compact radio sources are highly relativistic and Doppler boosted,
making studies of their intrinsic properties difficult.
Observed brightness temperatures can be used to study the intrinsic 
physical properties of the relativistic jets,
and constrain models of jet formation in the inner jet region.
We aim to observationally test such inner jet models.
The  very long baseline interferometry (VLBI)
cores of compact radio sources are optically 
thick at a given frequency. The distance of the core from 
the central engine is inversely proportional to the frequency.
Under the equipartition condition between the magnetic field 
energy and particle energy densities, the absolute distance of 
the VLBI core can be predicted. 
We compiled the brightness temperatures of VLBI cores 
at various radio frequencies of 2, 8, 15, and 86~GHz.
We derive the brightness temperature on sub-parsec scales
in the rest frame of the compact radio sources.
We find that the brightness temperature increases with increasing
distance from the central engine, indicating that the intrinsic jet
speed (the Lorentz-factor) increases along the jet.
This implies that the jets are accelerated in the (sub-)parsec
regions from the central engine.
\end{abstract}


\section{Introduction}

Relativistic jets are
frequently observed in many
compact astronomical objects such as 
gamma-ray bursts~(GRB)~\citep{kul+99,gre+03}, 
X-ray binaries~(XRB or microquasars)~\citep{MR94,MR98}, and
active galactic nuclei~(AGN)~\citep{cur18,BM54,BP84}.
Deep insights into the processes governing the generation, 
the physics, and the behavior of the relativistic jets can be obtained
by studying the differences or consistencies in the properties
of these objects. In an effort to understand these processes, 
the relativistic jets in AGN  have been 
extensively studied~\citep[see][]{KP81,BP84,zen97,fer98,HK06}. 

The apparent velocities of the relativistic jets in powerful AGN
show apparent superluminal motions~\citep{rees66,whi+71}
as high as 
$\sim50~c$~\citep[e.g.,][]{lis+09} 
in the radio regime, and rapid variability in the
flux and polarization of the relativistic jet component
on scales $\leq$ 1 pc. Although the jets may also contain 
non-relativistic jet components, there is evidence that the relativistic
component dominates and extends up to Mpc~\citep[e.g.,][]{wal+01}.
While the matter content of the relativistic jet remains uncertain,
there are indications that protons dominate the mass flux
even as electron-positron pairs dominate the particle flux
in relativistic AGN jets~\citep{SM00}. Magnetic fields are considered
the most likely driving mechanism on these scales as well~\citep{lov76,BZ77,BP82}.

In any case, the acceleration mechanism is not well understood in jets, 
with magnetohydrodynamic (MHD)
models~\citep[e.g.,][]{VK04,lyu09} currently being one of the favored scenarios.
MHD models are contested by models 
which attribute the mechanism to acceleration by 
magnetic fields~\citep[e.g.,][]{HB00,LR03,MB09,TT14,nok+15}.
In particular, \cite{MB09} demonstrated jet formation,
based on the three-dimensional, general relativistic MHD simulations
of rapidly rotating Kerr black holes, and showed that jets may be able to reach
up to a Lorentz factor of 10 depending on the field geometry.

The intrinsic properties of extragalactic jets are difficult to measure, 
owing to the relativistic motion of the jet plasma, which can result
in substantial 
time delays and Doppler boosting in the observer's frame. Time delays 
strongly affect optically thin emission along a given path through the jet, 
and Doppler boosting couples the true physical properties of the emitting 
material to the geometrical and kinematic conditions in the flow. 
We would like to derive the evolution of some intrinsic parameters of the flow
which can be used as a probe of the acceleration.
Brightness temperature measurements of VLBI cores as
derived from Very Long Baseline Interferometry (VLBI)
are such  a probe. 
It brings us the closest to the central 
engine, reduces the effect of time delays (important for the optically 
thin part of the flow), and also reduces the chance of contaminating 
our measurements due to variable Doppler boosting in a (likely) 
curved flow downstream of the core. 

Observations at 86~GHz are particularly well-suited for this work, 
since they penetrate closer to the origin of the flow, owing to 
reduced opacity at higher frequencies.
Sensitive VLBI observations at 86~GHz with large radio telescopes,
e.g., the Global Millimeter VLBI Array~\citep[GMVA, ][]{mar+12},
have been made for a few prominent AGN jets, such as
3C 111~\citep{Doeleman96}, 
3C 454.3~\citep{Krich95,Krich99,Pagels},
NRAO 150~\citep{Agudo+07,mol+14},
NRAO 530~\citep{Bower97},
M87~\citep{kri+06,had+16}, Mrk 501~\citep{gir+08,koy+16}, 
3C 273, 3C 279~\citep{att01,lee+15}, 
and S5 0716+714~\citep{ran+15}.
In order to further increase the number of objects,
which can be imaged at 86~GHz,
five detection and imaging surveys were conducted during the last few decades, 
with a total of 167 extragalactic radio sources 
observed~\citep[see][]{Bea97, Lonsdale98, Ranta98, Lobanov00, lee+08}.
In these surveys fringes were detected in 130 objects, 
and 114 radio sources have been successfully imaged.

In order to derive the intrinsic parameters of the relativistic jets,
it is possible to compare
the brightness temperatures measured at 86~GHz 
with those at lower frequencies (e.g., 2 -- 15~GHz).
As shown in \cite{lee13}, the brightness temperatures
of the VLBI cores imaged at 86~GHz are found to be lower than those
at 15~GHz, assuming
the brightness temperature on average is the same for all sources, and
the viewing angles of the target jets
are close to the critical value for the maximal apparent jet speed,
following \cite{hom+06}. Careful comparison of the brightness temperatures
at multifrequencies, that is, at 2, 8, 15, and 86~GHz
enabled the study
of the variation of the brightness temperatures as a function of frequency.
This suggests that the VLBI cores mark
the position of the transition zone
between magnetic and kinetic dominance~\citep{lee14}.

Under the equipartition condition between jet particle and magnetic 
field energy densities, the position shift of the VLBI cores between 
two frequencies can be predicted~\citep{lob98}. 
The brightness temperatures in the rest frame of sources and the 
predicted core shift can be used to test the inner jet models,
by deriving the intrinsic parameters of the relativistic jets.

In this paper, we investigate the evolution of
the brightness temperatures (and hence the Lorentz factor)
of the relativistic jets
by using the frequency dependence of the VLBI core
position.
In Section~2,
we describe a general strategy for compiling the brightness
temperatures and a model which was adopted to predict the physical
distance of the observed VLBI core from the central engine.  
In Section~3,
we present the evolution of the brightness temperature
resulting from the application of the predicted absolute distance
of VLBI cores to brightness temperatures in the source frame
on (sub-)parsec scales. 
In Section~4,
the intrinsic parameters of the relativistic jets are discussed,
and in Section~5 we provide conclusions.
Throughout the paper, we use a Hubble constant 
$H_{\rm 0} = 71\,{\rm km\,s}^{-1}\,{\rm Mpc}^{-1}$ and 
a cosmological density parameter $\Omega_{\rm m} = 0.27$. 
%Omega0=1 (q0=0.5*Omega0)
%OmegaL=0.73
%OmegaD=0.23
%OmegaB=0.04

\section{Brightness temperature in ultracompact jets}

\subsection{General strategy}\label{general}

In order to investigate the physics of compact jets in sub-parsec scale 
regions,
we combined core brightness temperature measurements
from 86~GHz~\citep{lee+08} with measurements obtained at lower frequencies,
namely at
2~GHz and 8~GHz~\citep{PK12}, 
% 5\,GHz~\citep{fom+00}, 
and 15~GHz~\citep{kov+05}. 
% and 43\,GHz~\citep{JM01}. 
\cite{lee+08} observed 127 compact radio jets using
the GMVA at 86.2~GHz in the period
between 2001 October and 2002 October, and imaged 109 jets
with a typical resolution of approximately $40~\mu$as
(i.e. a scale of $<0.1$ pc at $z=1$).
A 2-dimensional circular Gaussian model was fitted
to the VLBI core of each jet
for estimating a source-frame core brightness temperature $T_{\rm b}$
following
\begin{equation}
	\label{tb}
      T_{\rm b} =1.22\times10^{12} 
                   \frac{S}{d^2\nu^2}(1+z) \,\,{\rm K},
\end{equation}	
where $S$ is the fitted core flux density in Janskys,
$d$ is the angular size (or FWHM)
of the fitted component in milliarcseconds,
and $\nu$ is the observing frequency.
The factor $(1+z)$ reflects the cosmological effect on 
the brightness temperature.
A core component was considered to be unresolved
when its fitted FWHM was smaller than
its minimum resolvable size,
and hence the brightness temperature was determined to be a lower limit.
The minimum resolvable size of a component in a general VLBI image
is given by \cite{lob05} as
\begin{equation}
%d_{\rm min}=\frac{2^{1+\beta/2}}{\pi}{\left[{\pi ab\ln2\ln\left({\frac{SNR}{SNR-1}}\right)}\right]}^{1/2}, 
d_{\rm min}=2^{1+\beta/2}{\left[{\frac{ab\ln2}{\pi}\ln\left({\frac{SNR}{SNR-1}}\right)}\right]}^{1/2}, 
\label{eqn2}
\end{equation}
where {\it a} and {\it b} are the axes of the restoring beam of observations, 
{\it SNR} is the signal-to-noise ratio of the jet component,
and $\beta$ is a weighting function of imaging,
which is 0 for natural weighting
or 2 for uniform weighting.
If $d < d_{\rm min}$ for a core component, 
then the component is considered to be unresolved
and the lower limit of $T_{\rm b}$ is obtained with $d=d_{\rm min}$.

\cite{PK12} observed a sample of 370 compact radio jets
with up to 24 radio telescopes at 2.3~GHz and 8.6~GHz
in the period between 1998 October and 2003 September.
The VLBI cores fitted with the circular Gaussian model
had median sizes of 1.04 mas ($\sim 6.75$ pc)
and 0.28 mas ($\sim 1.90$ pc) at 2.3~GHz and 8.6~GHz, respectively,
and typical core brightness temperatures of $2.5\times10^{11}$ K at
both frequencies.
\cite{kov+05} observed 250 compact radio jets with the very long baseline
array (VLBA) at 15.4~GHz in the period between 1993 and 2003,
with the smallest jet angular size of 0.02-0.06 mas,
which corresponds to a linear scale of $<0.1$ pc for the nearest target.
The VLBI cores were fitted with the elliptical Gaussian model,
yielding brightness temperatures between $10^{11}-10^{13}$ K.

In selecting the core brightness temperatures of the compact radio jets
observed at multifrequencies in multiple epochs,
the measurements for the lower limits have been excluded.
From the multiple epoch 
measurements of the brightness temperatures at 15~GHz, 
we have taken their median value in order to take the near 
equipartition value~\citep{hom+06}.    
Fro the compilation of our brightness temperature data base
we included a source only when it was observed at more than two frequencies.
We found that a total of 109 out of 325 compact radio jets
were suitable for the sample as listed in Table~\ref{table:Lsyn}.

\subsection{Ultracompact jet and core shift}

Before we present our results, we want to discribe the theoretical
backgound.
We follow the standard relativistic jet model
of \cite{BK79},
who consider an idealized model of a steady radio jet,
in order to parametrize a relativistic jet. 
They assumed a narrow conical jet of small opening angle~$\phi$ 
whose axis makes an angle~$\theta$ with the line of sight of 
the observer
(the observed opening angle is $\phi_{\rm o} = \phi \csc \theta$). 
The jet is assumed to be supersonic and free, and to have a constant 
speed~$\beta_{\rm j}$. The magnetic field in the jet $B$ should vary 
as $r^{-1}$, where $r$ is the distance from the apex of the jet 
(harboring most likely the central engine). The flow of relativistic particles 
in the jet is accelerated by converting the internal relativistic 
particle energy~$\gamma_{\rm e}$ to 
the bulk kinetic energy~$\gamma_{\rm k}$.
Their particle energy distribution is 
$N(\gamma_{\rm e}) = N_0 \gamma_{\rm e}^{-s}$
for $\gamma_{\rm min}(r) < \gamma_{\rm e} < \gamma_{\rm max}(r)$, 
where $s$ is the particle energy power index. 
Those electrons radiate inhomogeneous synchrotron radiation with 
a spectral index $\alpha = (1 - s)/2$.
For a typical $\alpha = -0.5$ ($S_{\nu}\propto\nu^{\alpha}$),
the corresponding particle energy distribution is 
$N(\gamma_{\rm e}) = N_0 \gamma_{\rm e}^{-2}$.
Assuming the equipartition between the jet particle energy and the 
magnetic field energy, which is given by $k_{\rm e}\Lambda B^2/8\pi$ 
($k_{\rm e}\leq 1$ and $\Lambda = 
         {\rm ln}(\gamma_{\rm max}/\gamma_{\rm min})$),
the total radiated synchrotron power from the emission region extending 
from $r_{\rm min}$ to $r_{\rm max}$ in the jet is 
\begin{equation}
    \label{eqn:Lsyn}
%   L_{\rm syn} = 
%         \frac{1}{8} k_{\rm e} \Delta \gamma_{\rm j} 
%          \beta_{\rm j} c B^2 r^2 {\phi_{\rm o}}^2,
    L_{\rm syn} = 
          \frac{1}{8} k_{\rm e} \Delta \gamma_{\rm j}^2 
           \beta_{\rm j} c B^2 r^2 {\phi}^2,
\end{equation}
where $\Delta = {\rm ln}(r_{\rm max}/r_{\rm min})$.
    
The observed VLBI core at any given frequency is located in the region 
where the optical depth to synchrotron self--absorption is 
$\tau_{\rm s} = 1$.
Assuming that the magnetic field and particle density decrease with 
$r$ as $B = B_1 (r_1/r)^m$ and $N = N_1 (r_1/r)^n$, where $B_1$ and $N_1$ 
are the magnetic field and the electron density at $r_1 = 1\,{\rm pc}$,
respectively,
the corresponding~$\tau_{\rm s}$ is given by~\citep[see][]{RL79,lob98}:
\begin{equation}
    \label{eqn:tau}
    \tau_{\rm s}(r) = 
      C_2(\alpha) N_1 \left( \frac{eB_1}{2\pi m_{\rm e}}\right)^{\epsilon}
      \frac{\delta^{\epsilon}\phi_{\rm o}}
           {r^{(\epsilon m+n-1)} \nu^{\epsilon+1}},
\end{equation}
where $e$, $m_{\rm e}$ are the electron charge and mass, respectively, 
and $\delta$, $\phi_{\rm o}$ are the Doppler factor and the observed 
jet opening angle.
Here $\epsilon = 3/2 - \alpha$, and $C_2 (\alpha)$ is a constant 
at a given spectral index~\citep{BG70}. For a typical spectral index of
$\alpha = -0.5$, 
$C_2(\alpha) = 8.4 \times10^{10}$ in $cgs$ units. 
The physical distance of the observed VLBI core 
from the central engine is obtained by setting
the optical depth~$\tau_{\rm s}(r)$ to unity:
\begin{equation}
    \label{eqn:r}
    r = [ \nu^{-1} (1+z)^{-1} {B_1}^{k_{\rm b}} \{ 6.2\times10^{18} C_2(\alpha) 
    {\delta_{\rm j}}^\epsilon N_1 \phi_{\rm o}\}^{1/(\epsilon+1)} ]^{1/k_{\rm r}} {\rm pc},
\end{equation} 
where $k_{\rm r} = \{(3-2\alpha)m + 2n - 2 \}/(5 - 2\alpha)$  and 
$k_{\rm b} = (3 - 2\alpha)/(5 - 2\alpha)$. 
\cite{kon81} showed that it is most reasonable to use $m=1$ and $n=2$ 
to explain the observed X-ray and synchrotron emission from 
the ultra compact VLBI jets. 
In this case, the corresponding $k_{\rm r} = 1$
does not depend on the spectral index.
Then, we find a simple evolution of the magnetic field
decreasing with $r$ as $B = B_1 (r_1/r)$, and hence $Br=B_1r_1$=constant.
This model can be used to obtain $B_1$ taking into account
Equation (\ref{eqn:Lsyn}) as
\begin{equation}
    \label{eqn:B1}
    {B_1}=\left(L_{\rm syn} 
          \frac{8}{k_{\rm e} \Delta}
	  \frac{1}{\gamma_{\rm j}^2 \beta_{\rm j} c r_1^2 {\phi}^2}
	  \right)^{1/2}.
\end{equation}

By inserting Equation (\ref{eqn:B1}) into Equation (\ref{eqn:r}),
the absolute position of the observed VLBI core~$r$ is related 
with the total radiated synchrotron luminosity~$L_{\rm syn}$ as 
\begin{equation}
    \label{eqn:r2}
    r = \left[ \xi \cC_r L_{\rm syn}^{k_{\rm b}/2} \{ \nu (1+z) \}^{-1} \right]^{1/k_{\rm r}} {\rm pc},
\end{equation}
with  
\begin{equation}
    \label{eqn:r2-1}
%   \xi = 1.1\times10^{-37} \frac{8}{k_{\rm e}\Delta} \left[ 6.2\times10^{18} 
%   C_2 (\alpha)\right]^{1/k_{\rm r}(\epsilon + 1)}
    \xi = \left(\frac{8}{k_{\rm e}\Delta}\right)^{k_{\rm b}/2} \left[ 6.2\times10^{18} 
    C_2 (\alpha)\right]^{1/(\epsilon + 1)}
\end{equation}
and
\begin{equation}
    \label{eqn:r2-2}
%   \cC_r = \frac{ \left[ {B_1}^{k_{\rm b}}({\delta_{\rm j}}^{\epsilon}N_1 \phi_{\rm o})^{1/(\epsilon + 1)}\right]^{1/k_r}}
%              { \gamma_{\rm j} \beta_{\rm j} c {B}^2 {\phi_{\rm o}}^2},
    \cC_r = \frac{({\delta_{\rm j}}^{\epsilon}N_1 \phi_{\rm o})^{1/(\epsilon + 1)}}
               {({\gamma_{\rm j}^2 \beta_{\rm j} c r_1^2 {\phi}^2})^{k_{\rm b}/2}},
\end{equation}
where $L_{\rm syn}$ is in erg~s$^{-1}$ and $\nu$ is in Hz.

%%%%%%%%%%%%%%%%%%%%%%%%%%%%%%% FIGURE %%%%%%%%%%%%%%%%%%%%%%%%%%%%%
\begin{figure*}[!t]
\epsscale{1.6}
%\plotone{ALL.eps}
\plotone{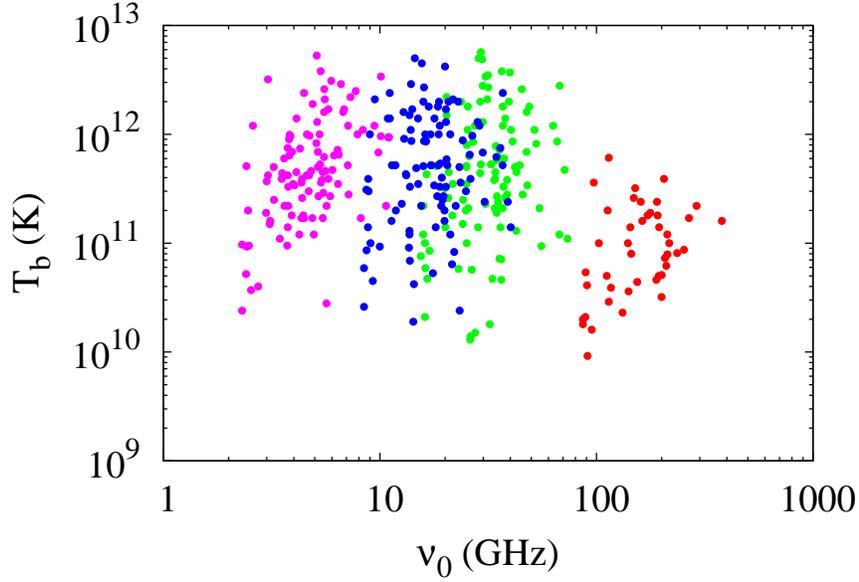}
\caption{Brightness temperatures in the source frame as a function 
             of frequency in the source frame.
             All available VLBI measurements (excluding lower limits) of 
             core components at 2~GHz (magenta dot), 8~GHz (blue dot),
		15~GHz (green dot), and 86~GHz (red dot) for
		the selected 109 sources. 
\label{fig1}}
\end{figure*}
%%%%%%%%%%%%%%%%%%%%%%%%%%%%%%% FIGURE %%%%%%%%%%%%%%%%%%%%%%%%%%%%%

%%%%%%%%%%%%%%%%%%%%%%%%%%%%%%% FIGURE %%%%%%%%%%%%%%%%%%%%%%%%%%%%%
\begin{figure*}[!t]
\epsscale{2.2}
\plotone{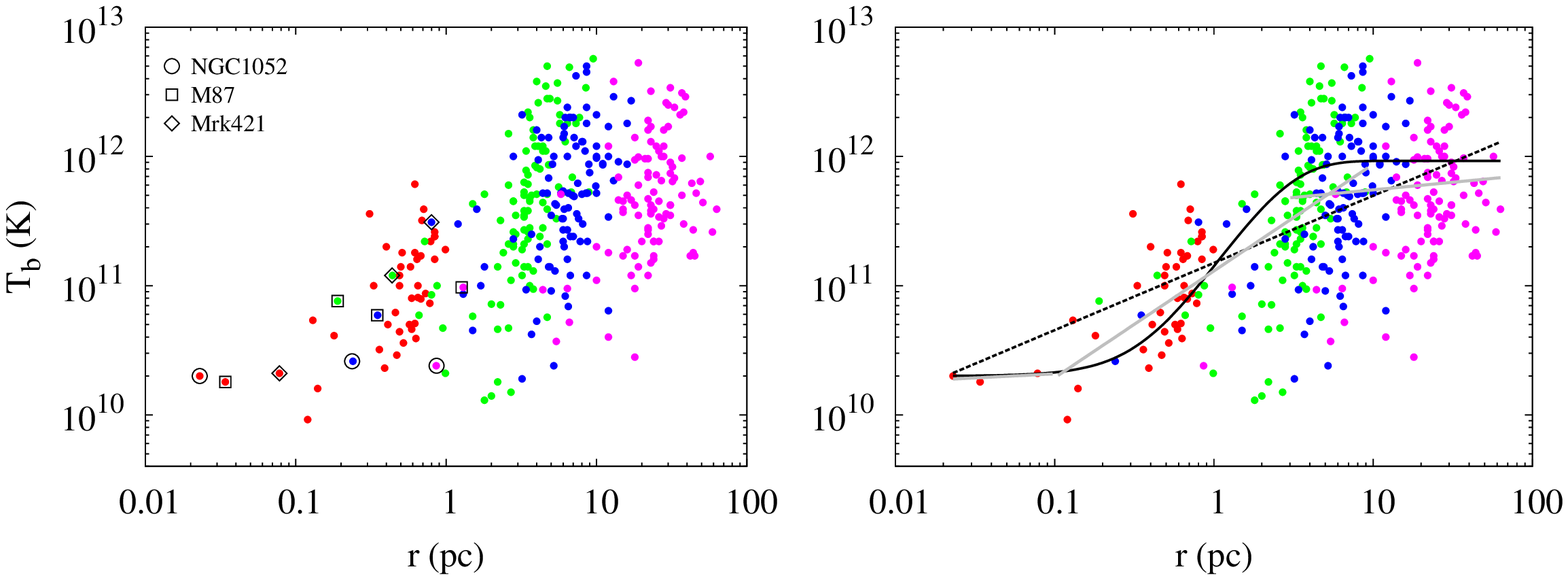}
\caption{[Left] Brightness temperatures in the source frame as a function 
             of the absolute position of the VLBI core components. 
             The used parameters for 
             the deduced position of the VLBI core are 
             $\gamma_{\rm j} = 10$,
             $\phi = 1/{\gamma_{\rm j}}^2$, 
             $N_1 = 5\times10^3\,{\rm cm}^{-3}$,
             and $r_{\rm max}/r_{\rm min} = 100$. 
	Colors are the same as for Figure~\ref{fig1}.
	Brightness temperatures for three sources are marked with
	symbols: circles for 0238$-$084 (NGC 1052),
	squares for 1228+126 (M87), and diamonds for 1101+384 (Mrk421).
	[Right] The same data with the best fits to the data
	for several cases, as described in Section~\ref{jetpara}:
	a single power law fit in black dashed line,
	multiple power law fits in grey solid line,
	and a fit of Equation~(\ref{eqn:model}) in black solid line.
\label{fig2}}
\end{figure*}
%%%%%%%%%%%%%%%%%%%%%%%%%%%%%%% FIGURE %%%%%%%%%%%%%%%%%%%%%%%%%%%%%

\section{Evolution of the brightness temperature}

In Figure~\ref{fig1} we combined the brightness temperature measurements
obtained at the different frequencies in one plot. We plot
the brightness temperature (in the source frame~$T_{\rm b}$)
as a function of frequency in the rest 
frame of the source, $\nu_{0} = \nu(1+z)$.
The brightness temperatures observed 
at 86~GHz are obviously lower than those at lower frequencies 
(2, 8, and 15~GHz).
In Table~\ref{table:Tbstat},
we quantify this result,
where we show for each frequency the median, mean
and standard deviation (rms) value of
the brightness temperature.
From the table it is seen that
the median and mean of
the 86~GHz brightness temperatures
are lower by a factor of 5-6 compared to
those of lower frequency brightness temperatures.
Despite the uncertainties due to 
the amplitude calibration error of $20\%-30\%$ at 86~GHz~\citep{lee+08} and 
the source variability of a factor of $\sim2$, the brightness temperatures 
observed at 86~GHz are still relatively lower than those
at the lower frequencies, implying that the decrease of brightness temperatures
at 86~GHz is significant.

The variation of brightness temperature with frequency
can be translated into a variation of brightness temperature
as a function of core separation
assuming the validity of the relativistic jet model,
as in particular using equation~(\ref{eqn:r}).
With the reasonable assumption of $m=1$, $n=2$,
we obtain $k_{\rm r}r=1$,
and therefore $r_{\rm core} \propto {\nu_{0}}^{-1/k_{\rm r}}$.
The decrease of the brightness temperature with frequency
(from 400 to 10 GHz in the source frame) therefore implies 
an increase of brightness temperature
with core separation by about one order of magnitude.
\begin{deluxetable}{lcccc}
\tabletypesize{\scriptsize}
\tablecaption{Statistics of brightness temperature\label{table:Tbstat}}
\tablewidth{0pt}
\tablehead{
\colhead{$\nu$ (GHz)} & 
\colhead{$T_{\rm b,med}$ (K)} & 
\colhead{$T_{\rm b,mean}$ (K)} & 
\colhead{$\sigma_{T_{\rm b}}$ (K)} &
\colhead{$N$}\\
\colhead{(1)} & 
\colhead{(2)} & 
\colhead{(3)} & 
\colhead{(4)} & 
\colhead{(5)}
}
\startdata
2.3 & 4.5e+11  & 8.0e+11 & 9.2e+11 & 104 \\ 
8.6 & 5.2e+11  & 8.5e+11 & 9.3e+11 & 104 \\ 
15.4& 4.5e+11  & 8.9e+11 & 1.1e+12 & 106 \\ 
86.2& 8.1e+10  & 1.3e+11 & 1.2e+11 & 44 \\ 
\enddata
\tablecomments{ 
Column designation:
(1)~-~observing frequency;  
(2)-(3)~-~median and mean brightness temperatures;
(4)~-~standard deviation of brightness temperature;
(5)~-~number of brightness temperature used.
}
\end{deluxetable}

However, the position of the core not only depends
on the source rest frame
frequency but also on the synchrotron luminosity of
the source (see equation~(\ref{eqn:r2})).
This may at least partly explain the relatively large scatter of
brightness temperature in Figure~\ref{fig1},
which indeed is based on many sources
with different synchrotron luminosity (see Table~\ref{table:Lsyn}). 
Another reason for the scatter may also come from source intrinsic deviations
from the idealized power law indices $m$ and $n$,
which may be slightly different for individual sources.
Since the particle energy and magnetic field energy densities 
at 1\,pc, $B_{1}$ and $N_{1}$, are different from source to source, 
and usually unknown, one cannot easily determine the absolute position 
of the core from equation~(\ref{eqn:r}) except for the case 
when the apparent shift of the core position at several frequencies is 
measured~\citep{lob98}.
However, with the assumption of energy equipartition between particle and
fields in a relaxed and steady jet,
the core position can be predicted with the known synchrotron 
luminosity~$L_{\rm syn}$ using equation~(\ref{eqn:r2}).   
The synchrotron luminosity can be estimated from the core flux 
measurements of each source over the range of rest frame frequencies, 
$\nu_{0,\rm min} \leq \nu_{0} \leq \nu_{0,\rm max}$. 
Fitting the spectrum of the core gives the total flux of the source 
over the range of frequencies.
Then, the synchrotron luminosity $L_{\rm syn}$ can be estimated
taking into account the total flux and luminosity distance of each source. 
%Then, the synchrotron luminosity $L_{\rm syn}$ is given by: 
%\begin{equation}
%    \label{eqn:Lsyn-2}
%    L_{\rm syn} = 4 \pi D_{\rm L}^2 F_{\rm t},
%\end{equation}
%where $D_{\rm L}$ is luminosity distance and 
%	$F_{\rm t}$ is the total flux of the source over the range of frequencies. 
   %and is given by:
   %\begin{equation}
   %\label{eqn:DL}
   %D_{\rm L} = \frac{c}{H_{\rm 0} q_{\rm 0}^2} ( z q_{\rm 0} + 
   %            ( q_{\rm 0} - 1)(\sqrt{2q_{\rm 0} z + 1} - 1),
   %\end{equation}
   %where $c$ is the speed of light and  $z$ is redshift. 
   %We use a Hubble constant $H_{\rm 0} = 100\,h\,{\rm km s}^{-1}\,{\rm Mpc}^{-1}$ and 
   %deceleration parameter $q_{\rm 0} = 0.5$. 
%   We use a Hubble constant 
%   $H_{\rm 0} = 100\,{\rm km s}^{-1}\,{\rm Mpc}^{-1}$ and 
%   a cosmological density parameter $\Omega_{\rm m} = 0.3$. 
%Table~\ref{table:Lsyn} lists the estimated synchrotron luminosity and 
%the core flux density measurements at four frequencies.
In Table~\ref{table:Lsyn} we summarize the estimated synchrotron luminosity,
which is used together with equation~(\ref{eqn:r2}) to drive the core separation.

Figure~\ref{fig2} shows the brightness temperatures 
as a function of the determined core position for all compact jets
selected. 
As described in Section~\ref{general},
all sources are assumed to have the same Lorentz factor 
$\gamma_{\rm j} = 10$, jet opening angle $\phi = 1/\gamma_{\rm j}^2$,
and viewing angle $\theta = 1/\gamma_{\rm j}$. 
%The magnetic field in the jets are assumed to be constant $B = 1$\,G, 
The electron density at 1\,pc is $N_1 = 5\times10^3\,{\rm cm}^{-3}$. 
The synchrotron emission is assumed to be emitted from the region with 
the scale factor of $r_{\rm max}/r_{\rm min} = 100$ at 
frequencies of $2\,{\rm GHz} \leq \nu_{0} \leq 400\,{\rm GHz}$.
After investigating the change of results depending on $\gamma_{\rm j}$,
we found that the distance $r$ of a core observed at a given frequency
mildly depends on $\gamma_{\rm j}$, as $r\propto\gamma_{\rm j}^{0.34}$,
yielding, for example, the ranges of $r=0.023-63$~pc for $\gamma_{\rm j}=10$
and of $r=0.050-14$~pc for $\gamma_{\rm j}=100$.
As expected from Figure~\ref{fig1}, the brightness temperatures 
increase from the inner region to the outer region of (sub-)parsec scales, 
which implies that the energy of the radiating particles increases
as they are driven out from the central engine
in the (sub-)parsec scale region of the jet.

\section{Discussion}\label{jetpara}

\subsection{Modeling the radial dependence of brightness temperature}

In order to interpret the increase of the brightness temperatures
along the jet on the scales of 0.01-100~pc,
we modeled it with a single power-law function
as shown in Figure~\ref{fig2}. One single power-law seems to reasonably
fit the data,
but in particular on sub-pc scales, a number of data points fall below
the best fit line. This may imply that a single power-law fit may not
be the best presentation of the radial dependence of brightness temperature. 
Although the number of data points in each bin of the distance
(e.g., 0.01-0.1~pc, 0.1-1~pc, 1-10~pc and 10-100~pc) is different each other,
the increasing pattern in the inner region (e.g., 0.01-0.1~pc) looks
a bit different from those in the outer region (0.1-10~pc).
Moreover the large scatter of data in 1-100~pc may make difficult
to find an increasing trend. Therefore, we tried to fit the data
with multiple power-law functions on restricted ranges of the distance
(0.01-0.1pc, 0.1-10pc, and 3-100pc). We found that the power index
of the best fit function ranged from 0.06 to 0.8,
as shown in Figure~\ref{fig2}. The multiple power-law functions indeed reveal
the slow increasing trends of the brightness temperature in the regions
of 0.01-0.1~pc and 3-100~pc, yielding power indices of 0.06 and 0.11, respectively. Therefore we concluded that the brightness temperature increases
with different slopes (or power indices) in different regions for our data.
Here caution should be taken for the fit results on the 0.01-0.1~pc regions,
since there are only three data points
(although those are all data we have for the region).

This result stimulated us to developing a single function
suitable to model the overall evolution of the brightness temperatures
along the jet. One of the best empirical models which we found
has the following form:
%
%\begin{equation}
%    \label{eqn:model}
%    T_{\rm b} = T_{0}\left(\frac{T_{\rm m}}{T_{0}}\right)^{1-(r~{\rm csch}(r))^a},
%\end{equation}
\begin{equation}
    \label{eqn:model}
    %T_{\rm b} = T_{0}+(T_{\rm m}-T_{0})\left[1-\left\{\frac{r}{{\rm sinh}r}\right\}^a\right],
    T_{\rm b} = T_{0}+(T_{\rm m}-T_{0})\{1-(r~{\rm csch}r)^a\},
\end{equation}
where $T_{0}$, $T_{\rm m}$ are the minimum (or initial)
and maximum brightness temperatures in K, respectively, and
$r$ is the distance of the VLBI core components
from the central engine in pc.
This model assumes that $T_{0}$ is the brightness temperature
near the jet base and $T_{\rm m}$ that in the downstream of the jet. 
We used $T_{0}=2.0\times10^{10}$~K which is corresponding to the mean of
the observed brightness temperatures at $r=0.02-0.1$~pc in Figure~\ref{fig2}.
The best fitting was obtained
for
$T_{\rm m}=(9.2\pm0.7)\times10^{11}$~K
and $a=0.9\pm0.3$,
as shown in Figure~\ref{fig2}.
This implies that the brightness temperatures on sub-parsec scales
are close to or even below the equipartition temperature of $5\times10^{10}$~K
(indicating that the jets may be magnetically dominated)
and start to increase on (sub-)parsec regions, reaching the inverse
Compton limit of $\sim10^{12}$~K on parsec scales.
Based on this best fitting model,
we now can discuss specific jet parameters,
like the the Doppler-factor and Lorentz factor.
This is done in the next two sections.

\subsection{Doppler factor}

One of the possible and very simple ways to understand
the variation of the brightness temperatures
as a function of the distance is
to assume that all of the observed variation in brightness temperatures
is caused by a variation of the Doppler factor.
Since the compact radio jets are highly relativistic,
their synchrotron emission is Doppler boosted and the boosting effect
is parametrized by the Doppler factor.
The Doppler factor is related to the physical properties
of the relativistic jets such as 
the Lorentz factor $\gamma_{\rm j}$,
the intrinsic brightness temperature $T_{0}$,
and the jet viewing angle $\theta_{\rm j}$:
\begin{equation}
    \label{eqn:doppler}
    \delta = \frac{1}{\gamma_{\rm j}(1-\beta {\rm cos}\theta_{\rm j})},
\end{equation}
\begin{equation}
    \label{eqn:doppler-Tb}
    T_{\rm b}=T_{0}\delta,
\end{equation}
where $\beta=(1-\gamma_{\rm j}^{-2})^{1/2}$
is the speed of the jet in the rest frame of the source in units of $c$.

\begin{figure}[!t]
\epsscale{1.0}
\plotone{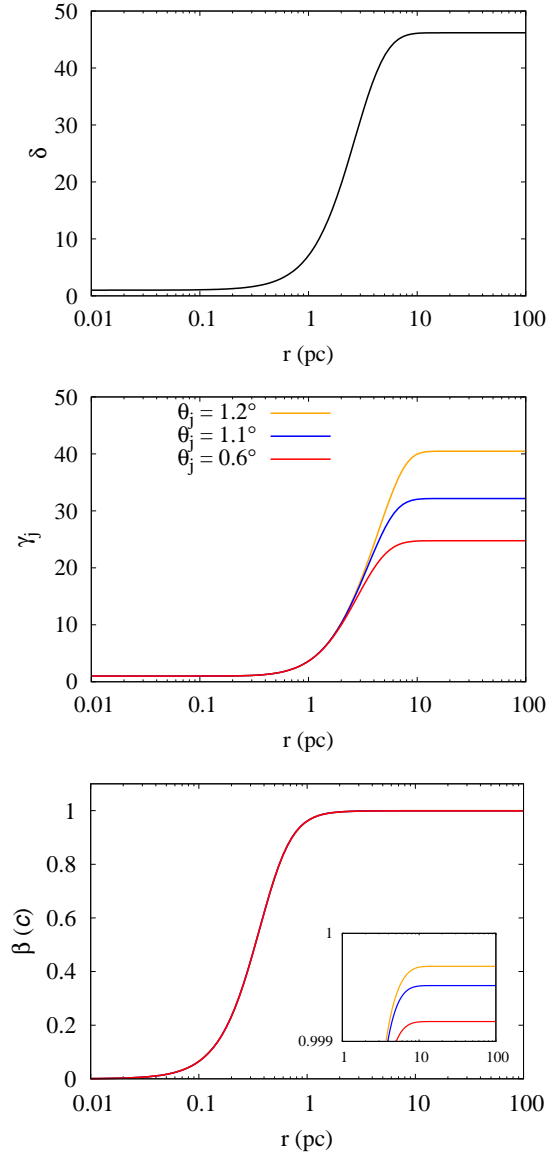}
\caption{Plots of Doppler factor (top panel), Lorentz factor (middle),
	and the jet speed (bottom), with a zoom-in inset,
	as a function of the absolution position of
             the VLBI core components. The colored lines are
		for the jet viewing angles of
		$\theta_{\rm j} = 1.2^{\circ}$ (orange),
		$1.1^{\circ}$ (blue),
		and
		$0.6^{\circ}$ (red).
             The initial Doppler factor is $\delta_{0}=1$. 
             %$\alpha = 0$. 
\label{fig3}}
\end{figure}

In order to derive the evolution of the Doppler factor as a function of
the distance $r$, equation~(\ref{eqn:doppler-Tb}) can be rewritten
as
\begin{equation}
    \label{eqn:doppler-Tb2}
    T_{\rm b}=T_{0}\left(\frac{\delta(r)}{\delta_{0}}\right),
\end{equation}
where $\delta_{0}$ is the Doppler factor at $r=r_{\rm min}$.
From equations~(\ref{eqn:model}) and (\ref{eqn:doppler-Tb}),
the Doppler factor $\delta(r)$ can be derived as
%
%\begin{equation}
%    \label{eqn:doppler-r}
%    \delta(r) = \delta_{0}\left(\frac{T_{\rm m}}{T_{0}}\right)^{1-(r~{\rm csch}(r))^a},
%\end{equation}
\begin{equation}
    \label{eqn:doppler-r}
    \delta(r) = \delta_{0}\left[1+\left(\frac{T_{\rm m}}{T_{0}}-1\right)\left\{1-(r~{\rm csch}r)^a\right\}\right],
\end{equation}
where we assume that the minimum (or initial) brightness temperature
in the model (equation~(\ref{eqn:model}))
is equal to the intrinsic brightness temperature.
With this assumption, the inferred Doppler factor
increases from
%$\delta_{0}=1$ to $\delta_{\rm m}\sim9.5$
$\delta_{0}=1$ to $\delta_{\rm m}\sim46$
in a similar way
as the brightness temperature does, as shown in Figure~\ref{fig3} (top panel).
%delta_m ~ 9.5
The maximum Doppler factor $\delta_{\rm m}$ constrains viewing angle
$\theta_{\rm j}$ in deducing the Lorentz factor as a function of the distance
from the central engine, as below. 

%%%%%%%%%%%%%%%%%%%%%%%%%%%%%%% FIGURE %%%%%%%%%%%%%%%%%%%%%%%%%%%%%
%%%%%%%%%%%%%%%%%%%%%%%%%%%%%%% FIGURE %%%%%%%%%%%%%%%%%%%%%%%%%%%%%

\subsection{Lorentz factor}

One of the intrinsic properties of the relativistic jets, the Lorentz factor,
can be deduced from equation~(\ref{eqn:doppler}) which, can be rewritten
as
\begin{equation}
    \label{eqn:gamma1}
    \delta^{2}{{\rm sin}^{2}\theta_{\rm j}}\gamma_{\rm j}^{2}-2\delta\gamma_{\rm j}+(1+\delta^{2}{{\rm cos}^{2}\theta_{\rm j}}) = 0.
\end{equation}
For $\delta$ and $\theta_{\rm j}$ satisfying the condition
$\delta^{2}{\rm sin}^{2}\theta_{\rm j}\leq1$,
a physically permitted solution of equation~(\ref{eqn:gamma1}) is
\begin{equation}
    \label{eqn:gamma2}
    \gamma_{\rm j} = \frac{1-{\rm cos}\theta_{\rm j} \sqrt{1-\delta^{2}{\rm sin}^{2}\theta_{\rm j}}}{\delta{\rm sin}^{2}\theta_{\rm j}}.
\end{equation}
By inserting equation~(\ref{eqn:doppler-r}) into equation~(\ref{eqn:gamma2}),
the dependence of the Lorentz factor on the distance
from the central engine $\gamma_{\rm j}(r)$
is deduced as shown in Figure~\ref{fig3} (middle panel),
and according to the definition $\beta=(1-\gamma_{\rm j}^{-2})^{1/2}$
the evolution of the jet speed $\beta(r)$ is directly deduced as shown
in Figure~\ref{fig3} (bottom panel).
Here it should be noted that $\theta_{\rm j}\le1/\delta_{\rm m}$.
Thus three examples of $\gamma_{\rm j}$ for
%$\theta_{\rm j}=3.0^{\circ}$, $5.0^{\circ}$, and $6.0^{\circ}$ are shown.
$\theta_{\rm j}=0.6^{\circ}$, $1.1^{\circ}$, and $1.2^{\circ}$ are shown.
This picture matches very well with similar acceleration plots given for
the magnetically driven, accelerating 
jet model~\citep[e.g.,][]{VK04,lyu09}.
\cite{VK04} argue that in the sub-parsec scale  
the mass flux is initially constant and then increasing, and in the outer region
the mass flux becomes constant again. The Lorentz factor also shows 
a similar trend, being constant 
in the inner region and then increasing in the outer region. As the mass flux
gets constant, the Lorentz factor also becomes constant in the outer region. 
We like to emphasize that such acceleration from sub-pc to pc-scales
has been reported in a number of cases, such as
NGC~6251~\citep{sud+00} 
with moderately relativistic speeds ($0.13c \sim 0.42c$),
Cygnus~A~\citep{bach+05,boc+16}
for apparent relativistic speeds ($0.2 \sim 0.5\,h^{-1}c$),
NGC~315~\citep{cot+99}
for parsec-scale jet acceleration ($0.75c \sim 0.95c$),
M~87~\citep{asa+14} for jet acceleration 
from non-relativistic ($0.01c$) to relativistic ($0.97c$),
and 3C\,345~\citep{unwin+97,LZ99} for highly relativistic speeds 
($\gamma_{\infty} \approx 35$).     
%NGC315 (Cotton et al. 1999):0.75c-0.95c over 3.4pc-9.5pc,
%or 1.1 c to 2.5 c at 5–20 × 10^3 rs (in Asada et al. 2013)
%
% CygA (Krichbaum et al. 1998): 0.2 c to 0.7 c at 2–10 × 10^3 rs 
%
% M87 (Asada et al. 2013): 0.01c to 0.97c over 200 rs - 10^5(?) rs

\subsection{Opacity effects and core blending}

The increase of core brightness temperature with decreasing frequency
could also be at least partly due to an instrumental beam blending effect,
with a larger observing beam size at lower frequencies.
Therefore the fitted core region may be dilluted by jet emission,
which comes from regions downstream of
the $\tau=1$ surface~\citep[see e.g.,][]{had+13}.
This blending effect will be more pronounced at the lower frequencies
(2, 8, 15 GHz), because of the larger beam sizes.

The brightness temperature measurements at 2.3~GHz and 8.6~GHz are based on
\cite{PK12}. Using a part of the database, \cite{kov+08} analyzed
opacity effect (or core-shift effect) on the VLBI core measurements.
They found that the median of the measured core shift between 2.3~GHz
and 8.6~GHz is 0.44 mas with its maximum of 1.4 mas. They discussed
the blending effect of the core and optically thin jet on the core shift
measurements.
They found that the blending shift of the core position
has a distribution with its median value of
0.006 mas \citep[see Figure 5 in][]{kov+08}.

Based on this study we may be able to estimate the blending effect
on the core flux measurement. \cite{PK12} found that the VLBI cores
fitted with the circular Gaussian model had median values of their sizes
of 1.04 mas and 0.28 mas at 2.3~GHz and 8.6~GHz, respectively.
These results, in addition to the median value of core-shift of 0.44 mas,
lead to a simple conical geometry of a jet as $d=1.7r$
($\frac{\delta d}{\delta r}=\frac{1.04-0.28}{0.44}=1.7$),
where $d$ is the jet size and
$r$ is the distance from the jet base,
assuming $r\propto \nu^{-1}$. Here, we may assume that the blending effect
of 0.006~mas obtained from the 8.6~GHz data is similar to
that for the 2.3~GHz data.
Then the true position of the 2.3~GHz VLBI core,
if it is observed with the beam of 8.6 GHz, would be shifted by 0.006~mas
to the jet base, compared with the observed position of the 2.3~GHz VLBI core.
According to the simple jet geometry,
the true (or blending-free) size of the core may decrease by
$\Delta d = 1.7 \Delta r \sim 1.7 \times 0.006$~mas = 0.01~mas
which is $~1\%$ of its size of 1.04 mas obtained with the model fitting.
If we assume that the $\tau=1$ surface of the core has uniform brightness
then the flux density of the true core may decrease by $<1\%$.
This implies that the blending effect to the core flux density measurement
may not be significant for the 2.3~GHz data with respect to the 8.6~GHz data.

We can also estimate the effect between the 8~GHz data and the 86~GHz data.
\cite{kov+08} also predicted that the expected core shift
between 8.6~GHz and an optical wavelength of 6000 Angstrom
would reach 0.12~mas which may be considered as the upper limit
of the core shift between 8.6~GHz and 86~GHz~\citep[e.g.,][]{had+11}.
\cite{lee+08} found
that the 86~GHz VLBI cores fitted with the circular Gaussian model
had median sizes of $\sim$0.04 mas.
These results again lead to a simple conical geometry of a jet as
$d=2r$ ($\frac{\delta d}{\delta r}=\frac{0.28-0.04}{0.12}=2$).
If we assume that the core blending effect
to the core shift is proportional to the frequency difference
(or resolution difference),
then the blending effect would be as large as $\sim$0.016~mas
($=\frac{86}{8.6}/\frac{8.6}{2.3}\times0.006$~mas) on the 8.6~GHz
data with respect to the 86 GHz data. As a result, the true size of the 8.6~GHz
core may decrease by $\Delta d =2\Delta r\sim 2\times 0.016$~mas$=0.032$~mas
which is $\sim13\%$ of the core size
of 0.24 mas as obtained at 8.6~GHz with model fitting.
Based on the constant brightness distribution of the $\tau=1$ surface of
the core at 8~GHz, the flux density $S_{\nu}$ of the true core may decrease
by $\sim24\%$. 
Then, we can estimate the fractional error ( or blending effect)
on the brightness temperature
$T_{\rm b}$ using Gaussian error propagation as
$\frac{\Delta T_{\rm b}}{T_{\rm b}}\approx \sqrt{(\frac{\Delta S_{\nu}}{S_{\nu}})^2+
4(\frac{\Delta d}{d})^2}=0.35$.
The effect may be significant but is still small compared to 
the observed difference in the brightness temperature between 8.6~GHz and 86~GHz
of a factor of 5-6 (see Table~\ref{table:Tbstat}).

\section{Conclusions}

Due to the highly relativistic physical environments, 
the compact jets of extragalactic radio sources are Doppler boosted,
making it impossible to directly measure their intrinsic properties:
the intrinsic brightness temperature, 
the Lorentz factor, 
and the viewing angle. 
Under the equipartition condition between the magnetic field energy 
and particle energy density, the absolute distance of the VLBI core
can be predicted, based on the facts that
(a) the VLBI cores of the compact radio sources are optically thick
at a given frequency,
and (b) the distance of the core from the central engine is
inversely proportional to the frequency.
The multifrequency large VLBI surveys of the compact radio jets
allowed the investigation of the source-frame brightness temperatures
of the (sub-)parsec regions of the relativistic jets,
yielding the following findings:
\begin{enumerate}
\item{
From the vicinity of the central engine to the outer region
of the relativistic jets on (sub-)parsec scales,
the brightness temperatures increase,
implying that the energy of the radiating particles increases
as they are driven out from the central engine
in the (sub-)parsec scale regions.
}
\item{
The evolution of the brightness temperatures on the (sub-)parsec scales
can be modeled, indicating that
the brightness temperatures on sub-parsec scales
are close to or even below
the equipartition temperature of $5\times10^{10}$~K
and start to increase on (sub-)parsec regions, reaching the inverse
Compton limit of $\sim10^{12}$~K on parsec scales.
}
\item{
The evolution of the brightness temperatures directly reveals
the dependence of the jet properties -- the Doppler factor, the Lorentz factor,
and the jet speed -- on the distance from the central engine
on (sub-)parsec scales, suggesting that
the relativistic jets are magnetically driven
to accelerate on the (sub-)parsec scale
and travel at a constant speed on the parsec scales.
}
\end{enumerate}

%

%%%%%%%%%%%%%%%%%%%%%%%%% Tables %%%%%%%%%%%%%%%%%%%%%%%%%
\begin{deluxetable}{lccccccccccc}
\tabletypesize{\scriptsize}
\tablecaption{Synchrotron luminosity\label{table:Lsyn}}
\tablewidth{0pt}
\tablehead{
\colhead{} & 
\colhead{} &
\colhead{$S_{\rm 86\,GHz}$} & 
\colhead{$T_{\rm b,86\,GHz}$} & 
\colhead{$S_{\rm 15\,GHz}$} & 
\colhead{$T_{\rm b,15\,GHz}$} & 
\colhead{$S_{\rm 8\,GHz}$} & 
\colhead{$T_{\rm b,8\,GHz}$} & 
\colhead{$S_{\rm 2\,GHz}$} & 
\colhead{$T_{\rm b,2\,GHz}$} & 
\colhead{$L_{\rm syn}$} &
\colhead{$B_{1}$}\\
\colhead{Name} & 
\colhead{$z$} & 
\colhead{(Jy)} & 
\colhead{(K)} & 
\colhead{(Jy)} & 
\colhead{(K)} & 
\colhead{(Jy)} & 
\colhead{(K)} & 
\colhead{(Jy)} & 
\colhead{(K)} & 
\colhead{(${\rm erg\,s}^{-1}$)} &
\colhead{(G)}
}
\startdata
0003$-$066 &     0.347 &   ... &    ... & 1.760 &2.6e+11 &  1.55 &1.6e+11 &  1.53 &1.5e+11 &2.4e+45 &  1.20 \\ 
0014$+$813 &     3.387 &   ... &    ... & 0.370 &1.2e+11 &  0.54 &5.2e+11 &  0.56 &9.6e+11 &6.1e+46 &  6.11 \\ 
0016$+$731 &     1.781 &   ... &    ... & 0.510 &4.5e+11 &  0.14 &2.4e+10 &  0.68 &3.5e+11 &1.1e+46 &  2.59 \\ 
0106$+$013 &     2.099 & 0.431 &1.7e+11 & 1.870 &1.6e+12 &   ... &    ... &  1.92 &1.2e+12 &3.7e+46 &  4.75 \\ 
0112$-$017 &     1.365 &   ... &    ... & 0.255 &4.6e+10 &  0.44 &1.6e+11 &  0.80 &1.7e+11 &3.3e+45 &  1.42 \\ 
0119$+$041 &     0.637 &   ... &    ... & 1.030 &1.0e+11 &  0.53 &6.9e+10 &  0.89 &2.2e+11 &4.5e+45 &  1.66 \\ 
0119$+$115 &      0.57 &   ... &    ... & 0.910 &1.5e+11 &  0.90 &4.3e+11 &  1.27 &6.0e+11 &2.2e+45 &  1.15 \\ 
0133$+$476 &     0.859 & 1.771 &2.4e+11 & 2.675 &5.0e+12 &  3.15 &4.5e+12 &  1.38 &7.4e+11 &1.5e+46 &  3.03 \\ 
0149$+$218 &      1.32 & 0.494 &5.1e+10 & 1.040 &7.2e+11 &   ... &    ... &  0.39 &3.6e+11 &1.2e+46 &  2.68 \\ 
0201$+$113 &     3.639 &   ... &    ... & 0.590 &4.7e+11 &  0.67 &2.4e+11 &  0.84 &2.2e+11 &1.1e+47 &  8.20 \\ 
0202$+$149 &     0.405 &   ... &    ... & 1.350 &1.8e+11 &  1.84 &5.2e+11 &  1.22 &5.0e+11 &9.0e+44 &  0.74 \\ 
0202$+$319 &     1.466 & 0.614 &7.9e+10 & 1.465 &1.4e+12 &  1.37 &2.0e+12 &  0.40 &2.2e+11 &1.9e+46 &  3.43 \\ 
0234$+$285 &     1.206 & 0.986 &2.4e+11 & 1.610 &2.1e+11 &  3.48 &1.8e+12 &   ... &    ... &2.5e+46 &  3.93 \\ 
0235$+$164 &      0.94 &   ... &    ... & 1.460 &2.8e+12 &  1.30 &8.7e+11 &  1.08 &2.4e+12 &1.9e+46 &  3.44 \\ 
0238$-$084 &     0.005 & 0.267 &2.0e+10 &   ... &    ... &  0.52 &2.6e+10 &  0.61 &2.4e+10 &5.0e+40 &  0.01 \\ 
0333$+$321 &     1.259 & 0.384 &1.4e+11 & 1.050 &2.7e+11 &  0.87 &5.4e+11 &  0.89 &2.7e+11 &9.1e+45 &  2.36 \\ 
0415$+$379 &    0.0491 & 1.104 &4.1e+10 & 0.970 &2.1e+10 &  0.45 &1.4e+11 &  0.70 &5.2e+10 &2.5e+43 &  0.12 \\ 
0430$+$052 &     0.033 & 1.107 &5.4e+10 & 1.565 &2.2e+11 &  0.84 &8.6e+10 &   ... &    ... &9.2e+42 &  0.08 \\ 
0440$-$003 &     0.845 &   ... &    ... & 0.645 &1.4e+11 &  0.45 &2.2e+11 &  0.63 &1.2e+11 &5.2e+45 &  1.78 \\ 
0521$-$365 &   0.05534 & 0.331 &9.2e+09 & 1.750 &5.9e+10 &  0.97 &3.0e+11 &  0.91 &9.3e+10 &7.8e+42 &  0.07 \\ 
0537$-$286 &     3.104 &   ... &    ... & 0.800 &1.2e+12 &  0.91 &6.2e+11 &  0.97 &1.2e+12 &1.1e+47 &  8.21 \\ 
0552$+$398 &     2.363 & 0.480 &2.2e+11 & 3.210 &1.1e+12 &  3.53 &1.3e+12 &  2.92 &2.5e+12 &7.6e+46 &  6.82 \\ 
0602$+$673 &      1.97 &   ... &    ... & 0.930 &2.8e+11 &  0.78 &3.0e+11 &  0.84 &1.7e+12 &5.1e+46 &  5.60 \\ 
0607$-$157 &    0.3226 & 0.965 &2.9e+10 & 4.050 &1.5e+12 &  3.19 &1.4e+12 &  1.58 &4.2e+11 &9.4e+44 &  0.76 \\ 
0615$+$820 &      0.71 &   ... &    ... & 0.285 &1.4e+10 &  0.34 &4.2e+10 &  0.52 &6.9e+11 &9.4e+44 &  0.76 \\ 
0642$+$449 &     3.396 & 0.590 &1.6e+11 & 2.490 &2.8e+12 &  3.19 &2.4e+12 &  0.84 &3.4e+12 &2.0e+47 & 11.04 \\ 
0716$+$714 &     0.127 & 1.048 &3.6e+11 &   ... &    ... &  0.85 &2.1e+12 &  0.41 &1.2e+12 &1.7e+44 &  0.33 \\ 
0727$-$115 &     1.591 &   ... &    ... & 2.870 &3.7e+12 &  3.40 &2.1e+12 &  2.50 &3.1e+12 &6.5e+46 &  6.30 \\ 
0736$+$017 &    0.1894 & 0.832 &1.0e+11 & 2.260 &5.1e+11 &  0.35 &9.3e+10 &  0.58 &4.0e+10 &2.4e+44 &  0.38 \\ 
0738$+$313 &     0.631 & 0.447 &3.6e+10 & 0.280 &3.5e+11 &  0.74 &1.3e+11 &  0.49 &1.4e+11 &2.4e+45 &  1.21 \\ 
0742$+$103 &     2.624 &   ... &    ... & 0.750 &9.4e+10 &  1.04 &2.4e+11 &  1.32 &1.1e+12 &6.1e+46 &  6.12 \\ 
0745$+$241 &     0.409 &   ... &    ... & 0.670 &3.5e+11 &  0.35 &2.0e+11 &  0.31 &2.5e+11 &2.3e+45 &  1.20 \\ 
0748$+$126 &     0.889 & 0.679 &1.6e+11 & 2.850 &2.0e+12 &  1.55 &2.0e+12 &  0.58 &1.7e+11 &7.2e+45 &  2.10 \\ 
0804$+$499 &     1.436 & 0.140 &6.2e+10 & 0.650 &6.0e+11 &  0.48 &5.2e+11 &  0.41 &4.6e+11 &5.6e+45 &  1.85 \\ 
0814$+$425 &     0.530 & 0.311 &2.3e+10 &   ... &    ... &  0.66 &1.6e+12 &  0.78 &3.9e+11 &8.8e+44 &  0.73 \\ 
0827$+$243 &     0.942 &   ... &    ... & 1.505 &1.3e+12 &  0.79 &8.6e+11 &  0.51 &1.7e+11 &4.0e+46 &  4.96 \\ 
0850$+$581 &    1.3173 & 0.104 &3.2e+10 & 0.060 &7.2e+10 &  0.31 &2.5e+11 &  0.40 &3.8e+12 &2.3e+45 &  1.19 \\ 
0851$+$202 &     0.306 & 0.618 &2.0e+11 & 1.260 &1.4e+11 &  1.95 &9.4e+11 &  1.60 &3.7e+11 &5.5e+44 &  0.58 \\ 
0859$+$470 &     1.462 & 0.222 &1.2e+11 & 0.510 &2.6e+11 &   ... &    ... &  0.35 &2.8e+10 &7.1e+45 &  2.09 \\ 
0917$+$449 &     2.180 &   ... &    ... & 0.865 &1.8e+12 &  0.52 &9.7e+11 &  0.25 &2.2e+12 &1.3e+47 &  9.08 \\ 
0923$+$392 &     0.698 &   ... &    ... & 0.210 &1.3e+10 &  0.33 &1.9e+10 &  0.49 &1.8e+11 &6.1e+44 &  0.61 \\ 
0945$+$408 &     1.249 & 0.363 &5.0e+10 & 0.770 &1.9e+11 &  0.76 &3.3e+11 &  1.55 &6.9e+11 &8.7e+45 &  2.30 \\ 
0955$+$476 &     1.873 &   ... &    ... & 1.680 &5.2e+11 &  1.66 &8.8e+11 &  1.09 &2.9e+12 &1.1e+47 &  8.17 \\ 
1049$+$215 &     1.300 &   ... &    ... & 0.950 &6.1e+11 &  0.33 &4.0e+11 &  1.02 &5.3e+11 &1.2e+46 &  2.73 \\ 
1055$+$018 &     0.888 &   ... &    ... & 1.370 &4.4e+11 &  1.08 &5.1e+11 &  1.24 &3.6e+11 &1.3e+46 &  2.80 \\ 
1101$+$384 &    0.0308 & 0.292 &2.1e+10 & 0.450 &1.2e+11 &  0.29 &3.1e+11 &   ... &    ... &2.1e+42 &  0.04 \\ 
1116$+$128 &     2.118 &   ... &    ... & 0.470 &3.4e+11 &  0.71 &3.9e+11 &  0.72 &2.8e+11 &2.5e+46 &  3.94 \\ 
1124$-$186 &     1.048 &   ... &    ... & 1.240 &3.5e+12 &  0.82 &5.2e+11 &  0.92 &4.2e+11 &1.9e+46 &  3.44 \\ 
1128$+$385 &     1.733 & 0.504 &8.1e+10 & 0.570 &5.0e+11 &  0.76 &2.0e+12 &   ... &    ... &2.3e+46 &  3.73 \\ 
1145$-$071 &     1.342 &   ... &    ... & 0.640 &1.4e+11 &  0.68 &2.7e+11 &  0.43 &3.8e+11 &1.2e+46 &  2.74 \\ 
\enddata
%\tablecomments{ 
%}
\end{deluxetable}
\setcounter{table}{0}
\begin{deluxetable}{lccccccccccc}
\tabletypesize{\scriptsize}
\tablecaption{Synchrotron luminosity (continued)}
\tablewidth{0pt}
\tablehead{
\colhead{} & 
\colhead{} &
\colhead{$S_{\rm 86\,GHz}$} & 
\colhead{$T_{\rm b,86\,GHz}$} & 
\colhead{$S_{\rm 15\,GHz}$} & 
\colhead{$T_{\rm b,15\,GHz}$} & 
\colhead{$S_{\rm 8\,GHz}$} & 
\colhead{$T_{\rm b,8\,GHz}$} & 
\colhead{$S_{\rm 2\,GHz}$} & 
\colhead{$T_{\rm b,2\,GHz}$} & 
\colhead{$L_{\rm syn}$} &
\colhead{$B_{1}$}\\
\colhead{Name} & 
\colhead{$z$} & 
\colhead{(Jy)} & 
\colhead{(K)} & 
\colhead{(Jy)} & 
\colhead{(K)} & 
\colhead{(Jy)} & 
\colhead{(K)} & 
\colhead{(Jy)} & 
\colhead{(K)} & 
\colhead{(${\rm erg\,s}^{-1}$)} &
\colhead{(G)}
}
\startdata
1150$+$812 &      1.25 &   ... &    ... & 1.055 &5.8e+11 &  0.68 &1.2e+12 &  0.41 &5.0e+11 &4.4e+46 &  5.21 \\ 
1156$+$295 &     0.725 & 1.629 &2.6e+11 & 0.910 &5.7e+10 &  2.42 &5.0e+12 &  1.22 &3.5e+11 &1.2e+46 &  2.71 \\ 
1219$+$285 &     0.103 & 0.186 &1.6e+10 & 0.280 &8.5e+10 &  0.16 &4.5e+10 &  0.18 &3.7e+10 &1.6e+43 &  0.10 \\ 
1228$+$126 &   0.00436 & 1.046 &1.8e+10 & 1.005 &7.6e+10 &  1.29 &5.9e+10 &  1.42 &9.7e+10 &1.6e+41 &  0.01 \\ 
1244$-$255 &     0.638 &   ... &    ... & 1.220 &8.0e+11 &  0.72 &3.3e+11 &  0.81 &3.3e+11 &7.3e+45 &  2.11 \\ 
1308$+$326 &     0.997 & 0.640 &1.8e+11 & 1.130 &2.3e+11 &   ... &    ... &  1.14 &4.2e+11 &7.7e+45 &  2.17 \\ 
1324$+$224 &      1.40 &   ... &    ... & 0.530 &2.1e+11 &  0.62 &3.3e+11 &  0.70 &1.6e+12 &1.1e+46 &  2.59 \\ 
1334$-$127 &     0.539 &   ... &    ... & 4.930 &2.0e+12 &  4.07 &9.1e+11 &  2.12 &4.4e+11 &3.7e+46 &  4.77 \\ 
1404$+$286 &     0.077 &   ... &    ... & 0.850 &4.7e+10 &  1.20 &1.0e+11 &  1.56 &9.5e+10 &2.5e+43 &  0.12 \\ 
1424$+$366 &     1.091 &   ... &    ... & 0.280 &1.8e+10 &  0.12 &5.3e+10 &  0.39 &4.6e+11 &2.1e+45 &  1.14 \\ 
1508$-$055 &     1.191 & 0.503 &4.6e+10 & 0.520 &3.8e+11 &  0.41 &2.7e+11 &   ... &    ... &8.5e+45 &  2.28 \\ 
1511$-$100 &     1.513 & 0.550 &1.0e+11 & 0.590 &2.8e+11 &  0.40 &1.2e+11 &  0.60 &3.9e+11 &1.8e+46 &  3.27 \\ 
1548$+$056 &     1.417 &   ... &    ... & 2.325 &3.3e+11 &  2.13 &5.9e+11 &  2.64 &2.1e+12 &5.1e+46 &  5.58 \\ 
1555$+$001 &     1.772 &   ... &    ... & 0.680 &1.3e+11 &  0.70 &5.0e+11 &  0.56 &7.2e+11 &2.6e+46 &  4.02 \\ 
1606$+$106 &     1.232 &   ... &    ... & 1.450 &7.8e+11 &  2.46 &2.0e+12 &  1.90 &1.3e+12 &9.7e+45 &  2.43 \\ 
1611$+$343 &     1.401 &   ... &    ... & 2.745 &1.8e+12 &  2.33 &1.7e+12 &  2.10 &6.2e+11 &8.2e+46 &  7.08 \\ 
1622$-$253 &     0.786 &   ... &    ... & 1.440 &1.2e+12 &  1.58 &1.4e+12 &  0.78 &4.5e+11 &7.5e+45 &  2.15 \\ 
1624$+$416 &      2.55 &   ... &    ... & 0.260 &2.1e+11 &  0.38 &6.8e+11 &  0.57 &1.7e+11 &1.8e+46 &  3.35 \\ 
1633$+$382 &     1.807 &   ... &    ... & 0.870 &5.1e+11 &  0.73 &3.6e+11 &  0.95 &6.5e+11 &3.3e+46 &  4.50 \\ 
1637$+$574 &     0.751 & 1.145 &3.2e+11 & 1.740 &8.5e+11 &  1.07 &4.9e+11 &  0.81 &3.4e+11 &7.0e+45 &  2.07 \\ 
1638$+$398 &     1.666 &   ... &    ... & 1.135 &8.6e+11 &  0.57 &2.2e+11 &  0.82 &6.4e+11 &4.7e+46 &  5.36 \\ 
1655$+$077 &     0.621 & 0.462 &1.0e+11 & 1.745 &2.0e+11 &  0.78 &9.1e+10 &  0.58 &9.5e+10 &2.0e+45 &  1.11 \\ 
1656$+$477 &     1.622 &   ... &    ... & 0.680 &1.1e+12 &  0.60 &8.3e+10 &  1.09 &4.7e+11 &1.6e+46 &  3.09 \\ 
1730$-$130 &     0.902 &   ... &    ... & 6.875 &5.7e+12 &  1.91 &2.7e+12 &  2.53 &3.9e+11 &1.3e+47 &  9.00 \\ 
1739$+$522 &     1.379 & 0.847 &3.9e+11 & 0.730 &3.8e+12 &  0.52 &4.2e+12 &  0.43 &2.9e+11 &1.9e+46 &  3.43 \\ 
1741$-$038 &     1.057 & 2.404 &1.9e+11 & 4.695 &2.7e+12 &  3.96 &5.2e+11 &  3.01 &9.7e+11 &3.4e+46 &  4.54 \\ 
1749$+$096 &     0.322 & 2.375 &6.1e+11 & 3.715 &2.2e+12 &  2.58 &2.4e+12 &  1.70 &3.2e+12 &2.2e+45 &  1.16 \\ 
1758$+$388 &     2.092 &   ... &    ... & 1.270 &6.9e+11 &  0.89 &6.5e+11 &  0.30 &5.2e+11 &2.3e+47 & 11.74 \\ 
1803$+$784 &    0.6797 & 0.785 &8.0e+10 & 1.555 &6.3e+11 &  1.52 &1.7e+12 &  1.27 &9.6e+11 &4.0e+45 &  1.56 \\ 
1807$+$698 &     0.051 &   ... &    ... & 0.640 &1.0e+11 &  0.36 &3.9e+11 &  0.49 &5.1e+11 &1.7e+43 &  0.10 \\ 
1821$+$107 &     1.364 &   ... &    ... & 0.270 &7.1e+10 &  0.39 &2.0e+11 &  0.73 &4.5e+11 &3.5e+45 &  1.47 \\ 
1823$+$568 &     0.664 & 0.485 &1.4e+11 & 0.850 &2.1e+11 &  0.75 &8.7e+11 &  0.40 &9.9e+11 &2.3e+45 &  1.17 \\ 
1849$+$670 &     0.657 &   ... &    ... & 1.610 &1.8e+12 &  0.91 &2.9e+12 &  0.33 &6.4e+11 &4.0e+46 &  4.95 \\ 
1908$-$201 &     1.119 &   ... &    ... & 2.600 &5.3e+11 &  0.99 &3.4e+11 &  1.17 &1.7e+11 &6.7e+46 &  6.42 \\ 
1921$-$293 &    0.3526 & 2.069 &3.9e+10 & 7.720 &4.5e+11 &  6.66 &5.2e+11 &  6.16 &1.6e+11 &2.5e+45 &  1.24 \\ 
1937$-$101 &     3.787 &   ... &    ... & 0.215 &1.1e+11 &  0.35 &1.4e+11 &  0.35 &9.4e+11 &4.7e+46 &  5.34 \\ 
1954$+$513 &     1.223 & 0.279 &1.8e+11 & 0.620 &4.5e+11 &  0.37 &1.0e+12 &  0.33 &5.3e+12 &5.7e+45 &  1.86 \\ 
1954$-$388 &     0.630 &   ... &    ... & 2.320 &5.0e+11 &  3.42 &1.5e+12 &  1.69 &7.5e+11 &3.6e+45 &  1.48 \\ 
1958$-$179 &     0.652 &   ... &    ... & 1.775 &1.1e+12 &  1.73 &1.1e+12 &  1.47 &9.0e+11 &9.8e+45 &  2.45 \\ 
2029$+$121 &     1.215 &   ... &    ... & 0.760 &3.9e+11 &  0.66 &5.2e+11 &  0.36 &8.3e+11 &2.6e+46 &  4.02 \\ 
2113$+$293 &     1.514 &   ... &    ... & 0.615 &3.8e+11 &  0.45 &1.0e+12 &  0.63 &1.7e+12 &1.6e+46 &  3.09 \\ 
2121$+$053 &     1.941 & 0.414 &8.7e+10 & 2.150 &2.6e+12 &   ... &    ... &  2.09 &1.6e+12 &4.0e+46 &  4.97 \\ 
2126$-$158 &      3.28 &   ... &    ... & 1.110 &8.6e+11 &  0.75 &7.5e+11 &  0.77 &6.8e+11 &2.1e+47 & 11.40 \\ 
2128$-$123 &     0.501 &   ... &    ... & 0.500 &5.8e+10 &  2.92 &2.3e+11 &  1.65 &1.1e+11 &2.8e+44 &  0.41 \\ 
2131$-$021 &     1.285 &   ... &    ... & 1.120 &2.4e+11 &  1.20 &2.2e+11 &  1.29 &1.0e+12 &2.1e+46 &  3.55 \\ 
2136$+$141 &     2.427 &   ... &    ... & 1.260 &8.2e+11 &  1.81 &1.2e+12 &  1.00 &1.0e+12 &5.9e+46 &  6.00 \\ 
2145$+$067 &     0.999 &   ... &    ... & 5.255 &3.4e+12 &  4.21 &1.8e+12 &  2.66 &1.0e+12 &1.1e+47 &  8.32 \\ 
2200$+$420 &    0.0686 &   ... &    ... & 1.870 &4.3e+11 &  1.22 &1.0e+12 &  1.42 &2.0e+11 &9.8e+43 &  0.24 \\ 
2201$+$315 &    0.2947 & 0.817 &5.0e+10 & 2.070 &3.2e+11 &  0.78 &1.4e+12 &  0.86 &1.9e+11 &6.2e+44 &  0.62 \\ 
2209$+$236 &     1.125 &   ... &    ... & 1.040 &5.3e+11 &  1.42 &1.4e+12 &  0.74 &1.9e+12 &7.6e+45 &  2.16 \\ 
\enddata
%\tablecomments{ 
%}
\end{deluxetable}
\setcounter{table}{0}
\begin{deluxetable}{lccccccccccc}
\tabletypesize{\scriptsize}
\tablecaption{Synchrotron luminosity (continued)}
\tablewidth{0pt}
\tablehead{
\colhead{} & 
\colhead{} &
\colhead{$S_{\rm 86\,GHz}$} & 
\colhead{$T_{\rm b,86\,GHz}$} & 
\colhead{$S_{\rm 15\,GHz}$} & 
\colhead{$T_{\rm b,15\,GHz}$} & 
\colhead{$S_{\rm 8\,GHz}$} & 
\colhead{$T_{\rm b,8\,GHz}$} & 
\colhead{$S_{\rm 2\,GHz}$} & 
\colhead{$T_{\rm b,2\,GHz}$} & 
\colhead{$L_{\rm syn}$} &
\colhead{$B_{1}$}\\
\colhead{Name} & 
\colhead{$z$} & 
\colhead{(Jy)} & 
\colhead{(K)} & 
\colhead{(Jy)} & 
\colhead{(K)} & 
\colhead{(Jy)} & 
\colhead{(K)} & 
\colhead{(Jy)} & 
\colhead{(K)} & 
\colhead{(${\rm erg\,s}^{-1}$)} &
\colhead{(G)}
}
\startdata
2216$-$038 &     0.901 &   ... &    ... & 2.025 &5.3e+11 &  1.43 &8.7e+11 &  0.37 &2.6e+11 &1.1e+47 &  8.15 \\ 
2223$-$052 &     1.404 & 0.642 &7.3e+10 & 3.740 &1.2e+12 &  3.49 &1.3e+12 &  2.39 &2.6e+12 &2.7e+46 &  4.04 \\ 
2227$-$088 &     1.562 &   ... &    ... & 2.030 &2.0e+12 &  0.43 &6.4e+10 &  0.77 &2.7e+11 &1.1e+47 &  8.13 \\ 
2230$+$114 &     1.037 &   ... &    ... & 2.680 &2.1e+12 &  1.74 &1.0e+12 &  2.34 &3.0e+11 &3.5e+46 &  4.64 \\ 
2234$+$282 &      0.79 & 0.405 &4.4e+10 & 0.210 &1.5e+10 &  1.22 &3.5e+11 &  1.80 &1.4e+12 &2.7e+45 &  1.28 \\ 
2243$-$123 &     0.630 &   ... &    ... & 1.920 &5.1e+11 &  1.11 &1.2e+11 &  1.38 &4.2e+11 &1.0e+46 &  2.49 \\ 
2255$-$282 &     0.927 &   ... &    ... & 5.650 &4.9e+12 &  5.99 &1.0e+12 &  0.89 &1.8e+11 &4.6e+46 &  5.33 \\ 
2329$-$162 &     1.153 &   ... &    ... & 0.610 &4.7e+10 &  0.93 &1.4e+11 &  0.91 &1.2e+11 &4.0e+45 &  1.57 \\ 
2345$-$167 &     0.576 &   ... &    ... & 1.215 &2.5e+11 &  1.52 &4.2e+11 &  0.92 &2.2e+11 &2.2e+45 &  1.17 \\ 
\enddata
%\tablecomments{ 
%}
\end{deluxetable}

\acknowledgments
%We gratefully thank the staff of the observatories participating in the GMVA;  
%the MPIfR Effelsberg 100-m telescope, the two IRAM telescopes on Plateau de Bure and
%Pico Veleta, the Mets\"ahovi Radio Observatory, the Onsala Space Observatory, and the VLBA.
We would like to thank the anonymous referee for important comments
and suggestions which have enormously improved the manuscript.
This research has made use of data obtained with the Global Millimeter VLBI Array (GMVA), which consists of telescopes operated by the MPIfR, IRAM, Onsala, Metsahovi, Yebes and the VLBA. The data were correlated at the correlator of the MPIfR in Bonn, Germany. The VLBA is an instrument of the National Radio Astronomy Observatory, a facility of the National Science Foundation operated under cooperative agreement by Associated Universities, Inc

%\appendix
%\section{Appendix material}

\end{document}